\documentclass[aps,prb,twocolumn,showpacs]{revtex4}
\usepackage{amsmath}

\usepackage{graphicx}

\begin{document}

\title{Magnetization relaxation, critical current density and vortex dynamics in a Ba$_{0.66}$K$_{0.32}$BiO$_{3+\delta}$ single crystal}

\author{Jian Tao, Qiang Deng, Huan Yang, Zhihe Wang, Xiyu Zhu, and Hai-Hu Wen}\email{hhwen@nju.edu.cn}

\affiliation{National Laboratory of Solid State Microstructures and Department of Physics, Collaborative Innovation Center of Advanced Microstructures, Nanjing University, Nanjing 210093, China}
\date{\today}

\begin{abstract}
We have conducted extensive investigations on the magnetization
and its dynamical relaxation on
a Ba$_{0.66}$K$_{0.32}$BiO$_{3+\delta}$ single crystal. It is
found that the magnetization relaxation rate is rather weak
compared with that in the cuprate superconductors, indicating a
higher collective vortex pinning potential (or activation energy),
although the intrinsic pinning
 potential $U_\mathrm{c}$ is weaker. Detailed analysis
leads to the following discoveries: (1) A second-peak effect
on the magnetization-hysteresis-loop was observed in a very wide
temperature region, ranging from 2K to 24K. Its general behavior
looks like that in YBa$_2$Cu$_3$O$_7$; (2) Associated with the
second peak effect, the magnetization relaxation rate is inversely
related to the transient superconducting current density $J_\mathrm{s}$
revealing a quite general and similar mechanism for the second
peak effect in many high temperature superconductors; (3) A
detailed analysis based on the collective creep model reveals a
large glassy exponent $\mu$ and a small intrinsic pinning
potential $U_\mathrm{c}$; (4) Investigation on the volume pinning force
density shows that the data can be scaled to the formula
$F_{p}\propto b^p(1-b)^q$ with $p=2.79$ and $q=3.14$, here $b$ is the reduced magnetic field to the irreversible magnetic field. The maximum normalized pinning force density appears near $b\approx0.47$. Finally, a vortex phase diagram is drawn
for showing the phase transitions or crossovers between different
vortex phases.
\end{abstract}


\maketitle

\section{Introduction}
Investigation on vortex physics is very important concerning the
potential high-power applications of a superconductor. In the
cuprate superconductors, due to the very high superconducting
transition temperature, layered structure, short coherence length,
strong thermal fluctuation etc., the vortex physics is extremely
rich, which has led to the unprecedented prosperous development on
the vortex physics\cite{BlatterReview}. Many new concepts and
phenomena, such as collective vortex creep\cite{Vinokur}, vortex
glass\cite{Fisher1,Fisher2,Koch}, first order vortex
transitions\cite{ZeldovNature,Kwok}, vortex
melting\cite{VortexmMelting}, second peak effect of
magnetization\cite{LingS,Bhattacharya} etc. have been proposed or
discovered. In the iron based superconductors, the vortex physics
looks quite similar to the
cuprate\cite{BSPRB81,vanderBeek,Konczykowski} although the
anisotropy of $\xi_{ab}/\xi_c$ is only about 2-5 which is much
smaller than that in the cuprate system\cite{JiaYAPL,WangZSPRB}.
Preliminary experimental studies have revealed that the vortex
dynamics in iron pnictide may be understood with the model of
thermally activated flux motion within the scenario of collective
vortex pinning\cite{YangHPRB,ProzorovPRB,ProzorovPhysicaC}. A
second-peak (SP) effect on the magnetization-hysteresis-loop (MHL)
has also been observed in
Ba$_{1-x}$K$_x$Fe$_2$As$_2$\cite{YangHAPL} and
Ba(Fe$_{1-x}$Co$_x)_2$As$_2$ single
crystals\cite{ProzorovPRB,ProzorovPhysicaC,ProzorovPRB2009,YNakajida}.
Beside the three typical high temperature superconductors, namely
the cuprate, MgB$_2$ and the iron based superconductors, the
Ba$_{1-x}$K$_x$BiO$_3$ \cite{JohnsonPRB,CavaNature332}(hereafter
abbreviated as BKBO) superconductor is quite special in terms of
its relatively high transition temperature (The highest
$T_\mathrm{c}$ can reach about 34 K\cite{JonesJSSC78}) and almost
three-dimensional feature\cite{SYPeiPRB,SchneemeyerNature}. In the
BKBO superconductors, the coherence length detected from scanning
tunneling microscope (STM) and other measurements is about 3-7
nm\cite{JangPRB,AffrontePRB,KwokPRB40,Schweinfurth}. The
structural characteristics\cite{SYPeiPRB}, the coherence length,
the Ginzburg-Landau parameter\cite{SNBariloPRB}
$\kappa=\lambda_\mathrm{L}/\xi$ seem to be very different from
those in the cuprate
superconductors\cite{VaugheyChemM,ShepelevPhysicaC,WorthingtonPRL,GallagherJAP}.
These peculiarities may bring about new ingredients to us in
understanding the vortex physics in high temperature
superconductors. Therefore we have grown the
Ba$_{1-x}$K$_x$BiO$_3$ single crystals and investigated the vortex
physics extensively with measurements of magnetization and its
dynamical relaxation.

\section{Experiment}

\begin{figure}
\includegraphics[width=8cm]{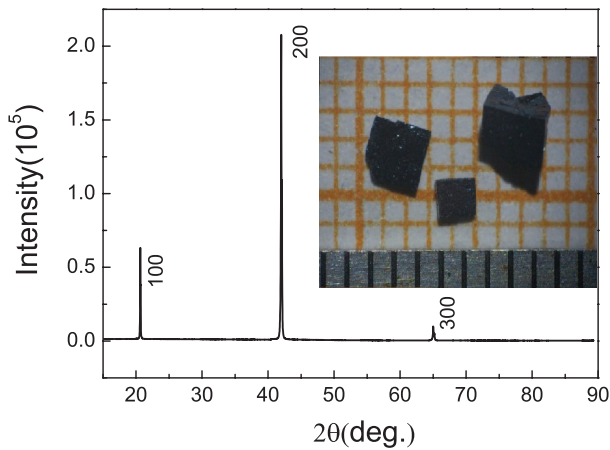}
\caption{(Color online) X-ray diffraction pattern of a crystal
with composition Ba$_{0.66}$K$_{0.32}$BiO$_{3+\delta}$. The
inset shows the photograph of BKBO crystals grown by
the electrochemical method. }\label{fig1} \end{figure}

The single crystals with high quality studied in this work were
prepared by the molten salt electrochemical method
presented previously\cite{NortonMRB,Nishio}. In the process of
electrochemical growth, the working electrode was made from 0.5
mm-diameter platinum wire (Alfa Aesar, 4N), and the working current
was 2 mA. In addition, we placed 43 g of KOH (J$\&$K Chemical
Ltd.) in a 100 cm$^3$ Telfon container, and heat it up to
250 $^\circ C$, staying for several hours until KOH was completely
melted, then added 1.49 g of Ba(OH)$_2$$\cdot$8H$_2$O (J$\&$K
Chemical Ltd., 2N) and 3.22 g of Bi$_2$O$_3$ to the
molten KOH solution, the growth begins after stirring the solution
for almost two hours. In this way, the crystals can be
successfully obtained with the size up to several millimeters if the
growing time is long enough. The best growth time in the experiment is around 48 hours.
Inset of Fig. 1 shows the photograph of samples we synthesized
through the electrochemical reaction method. By the way, all the
samples we measured were polished to a proper thickness in order
to guarantee the homogeneity. The lattice structure of the sample
was characterized by x-ray diffraction (XRD) at room temperature
with a Bruck-D8-type diffractometer. The XRD pattern of a sample is shown in Fig. 1, the vast value of the intensity
of the ($l$00) indices from the XRD pattern demonstrates the
$a$-axis orientation of the single crystal. The $a$-axis lattice
constant is 4.2995$\mathrm{\AA}$ through calculating the indexed peaks. The
sample composition was analyzed by using the energy dispersive x-ray
spectrometer (EDX/EDS). We concluded that the composition of
measured sample is about Ba$_{0.66}$K$_{0.32}$BiO$_{3+\delta}$,
where the oxygen content cannot be accurately determined.

\begin{figure}
\includegraphics[width=8cm]{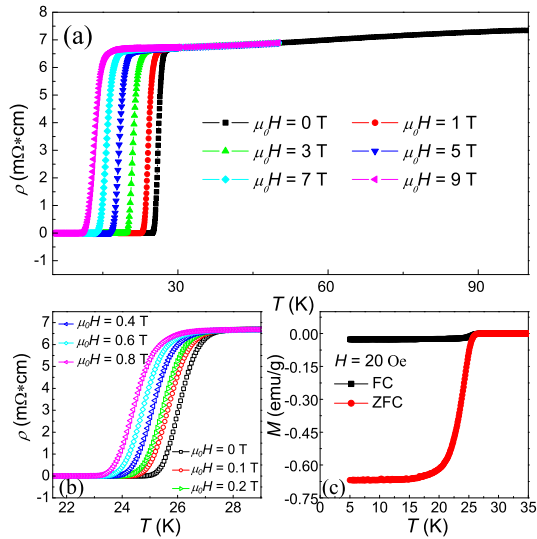}
\caption{(Color online)(a) Temperature dependence of resistivity
at different magnetic fields ranging from 0 T to 9 T. (b) Temperature
dependence of resistivity at different magnetic fields ranging
from 0 T to 0.8 T. (c) Temperature dependence of the diamagnetic
moment measured in processes of ZFC and FC at a magnetic field
of 20 Oe.}\label{fig2}
\end{figure}

The electric transport and magnetization measurement were
performed by a physical property measurement system (PPMS, Quantum
Design) and SQUID vibrating sample magnetometer (SQUID-VSM,
Quantum Design), respectively. Fig. 2 (a) and (b) show the
temperature dependence of resistivity for the crystal
Ba$_{0.66}$K$_{0.32}$BiO$_{3+\delta}$ under different magnetic
fields ranging from 0 T to 9 T. The onset transition temperature
at zero field is about 27 K by taking a criterion of 90$\%\rho_\mathrm{n}$,
here $\rho_\mathrm{n}$ is the normal state resistivity. The
diamagnetic moment of the sample is shown in Fig. 2(c) which was
measured in the zero-field-cooled (ZFC) and field-cooled (FC) mode
under a DC magnetic field of 20 Oe. The ZFC curve displays a
transition with an onset temperature around 26.5 K. The transition
temperature of the present sample is in
agreement with the phase diagram reported by other
group\cite{SamataJPCS}. From the results of XRD, resistivity and
diamagnetic measurements, the quality of the sample has been
proved to be good enough to do further study of the vortex
dynamics. In Fig. 3 we show the MHL curves with a magnetic field sweeping
rate $dB/dt$ of 200 Oe/s and 50 Oe/s at different temperatures ranging
from 2 K to 24 K (the magnetic field is
vertical to the $ab$ plane of the sample). The symmetric MHL curves demonstrate that the
measured sample is bulk superconductive and the vortex pinning is
bulk in nature. Undoubtedly, the dia-magnetization here is not due to
the surface shielding effect, and the Bean critical state model
can be used.

\begin{figure}
\includegraphics[width=8cm]{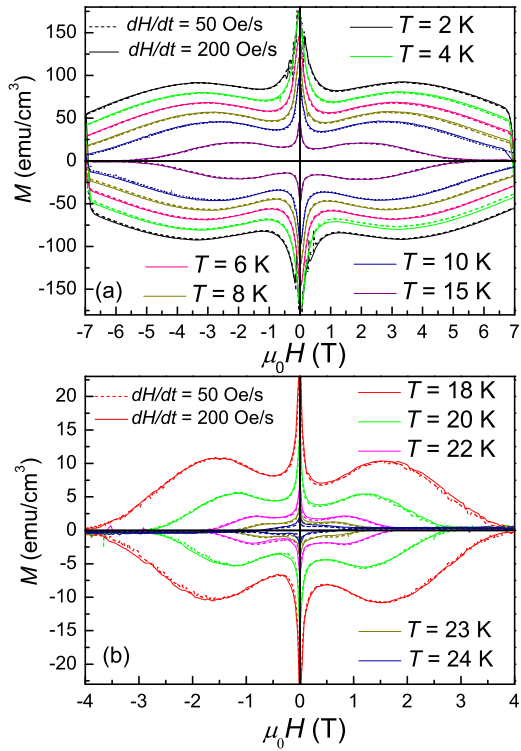}
\caption{(Color online) Magnetization hysteresis loops of the
BKBO single crystal at various
temperatures ranging from 2 K to 15 K (a), 18 K to 24 K (b). The
solid lines stand for the magnetization measured with field sweeping
rate of 200 Oe/s while the dash lines represent those measured with 50 Oe/s. We need to note that the magnetization at 4 K in the field ascending process with 50 Oe/s experienced a small flux jump at a low field. This flux jump keeps giving influence on the MHL at high field on this curve. For calculating the $\Delta M$, $J_s$ and $Q$ for 4K, we used the data in the left-hand side and second quadrant with negative magnetic field at 4K.}\label{fig3}
\end{figure}

\section{Models and analysis method}
\subsection{Thermally activated flux motion and collective flux creep}
To fully understand the vortex motion in
the BKBO single crystal, we start
from the model of thermally activated flux
motion\cite{AndersonPRL}:
\begin{equation}
E=v_0B \exp (-\frac{U(J_\mathrm{s},T,B_\mathrm{e})}{k_\mathrm{B}T}).
\end{equation}
Here $E$ is the electric field induced by the vortex motion, $v_0$
is the attempting moving velocity of the hopping vortex lines,
$U(J_\mathrm{s}, T, B_\mathrm{e})$ is the effective activation
energy, and $B_\mathrm{e}$ is the external magnetic field, $B$ is
the local averaged magnetic induction. Based on the vortex glass
\cite{Fisher1} and the collective pinning models\cite{Vinokur}, it
is predicted that $U(J_\mathrm{s},T,B_\mathrm{e})$ is positively
related to $[J_\mathrm{c}(T,B_\mathrm{e})/J_\mathrm{s}]^\mu$,
where $\mu$ is the glassy exponent describing the activation
energy. In order to ensure that $U(J_\mathrm{s},T,B_\mathrm{e})$
reaches zero when the external applied current $J_\mathrm{s}$
approaches the critical current $J_\mathrm{c}$, then Malozemoff
proposed to rewrite the activation energy in a very general way
as\cite{MalozemoffPHYSICAC}
\begin{equation}
U(J_\mathrm{s},T,B_\mathrm{e})=\frac{U_\mathrm{c}(T,B_\mathrm{e})}{\mu(T,B_\mathrm{e})}[(\frac{J_\mathrm{c}(T,B_\mathrm{e})}{J_\mathrm{s}(T,B_\mathrm{e})})^{\mu(T,B_\mathrm{e})}-1]
\end{equation}
where $U_\mathrm{c}$ and $J_\mathrm{c}$ are the  characteristic (or called as the
intrinsic) pinning
energy and initial critical current density (unrelaxed),
respectively. The glassy exponent $\mu$ gives different values for
different regimes of flux creep. From the elastic manifold theory
\cite{Vinokur}, it is predicted that $\mu = 1/7$, 3/2, and 7/9 for
the single vortex, small bundles, and large bundles of vortex
motion respectively. Interestingly, Eq. (2) has actually covered
many models describing the $U(J_\mathrm{s})$. For instance, when $\mu= -1$,
it will go back to the Kim-Anderson model\cite{AndersonPRL}, and
$\mu= 0$ corresponds to the Zeldov's logarithmic
model\cite{ZeldovAPL}. Furthermore, when the $J_\mathrm{c}$ is much larger
than $J_\mathrm{s}$, Eq. (2) will return to the form of collective pinning
models. Therefore, the value of $\mu$ will play a significant role
in understanding the vortex motion.

\subsection{Models for analyzing the magnetization relaxation}
For the sake of discussion, we will calculate the transient current density $J_\mathrm{s}$
from the width $\Delta M$ of MHLs, where $\Delta M = M^{-}-M^{+}$ with $M^{-}$($M^{+}$) the magnetization at a certain magnetic field in the increasing (decreasing)-field process. According to the Bean critical
state model\cite{BEANRMP36}, the transient superconducting
current density $J_\mathrm{s}$ can be expressed as
\begin{equation}
J_\mathrm{s}=20\frac{\Delta M}{w(1-\frac{w}{3l})},
\end{equation}
where the unit of $\Delta M$ is emu/cm$^3$, $w$, $l$ are the width
and length of the sample measured in cm ($w<l$), respectively.
In this work, we utilized the dynamical relaxation method to study
the vortex dynamics, instead of using the conventional relaxation
method\cite{MJPRB,WenHHPhysicaC95,WENPRB52}. The corresponding
physical quantity is the magnetization-relaxation rate $Q$ which is
defined as:
\begin{equation}
Q \equiv \frac{\textrm{d} \ln J_\mathrm{s}}{\textrm{d} \ln (dB/dt)}=
\frac{\textrm{d} \ln (\Delta M)}{\textrm{d}
\ln(\textrm{d}B/\textrm{d}t)}.
\end{equation}
The dynamical relaxation measurements are followed in this
way: the sample is cooled down to a certain temperature at ambient
magnetic field, and then we measure the MHL curves with two
different magnetic field sweeping rates.

From the general formulas Eqs. (1) and (2) mentioned above and the
definition of $Q$, we will employ the following expression to
quantify the characteristic pinning energy as derived by Wen et
al.\cite{WENJAC}
\begin{equation}
\frac{T}{Q(T,B_\mathrm{e})}=\frac{U_\mathrm{c}(T,B_\mathrm{e})}{k_B}+\mu(T,B_\mathrm{e})CT,
\end{equation}
here $C=\ln(2\nu_0B/(ldB_\mathrm{e}/dt)$) is a parameter which is weakly
temperature dependent, $l$ is the lateral dimension of the sample.
According to Eq.(5), in the low temperature region, the term of
$\mu(T,B_\mathrm{e})CT$ is much smaller than the term of $U_\mathrm{c}(T,B_\mathrm{e})/k_B$,
so we could ignore $\mu(T,B_\mathrm{e})CT$, and get $T/Q(T,B_\mathrm{e})\approx
U_\mathrm{c}(T,B_\mathrm{e})/k_B $, and $Q$ should show a linear dependence on $T$.
However, as we will show below, in BKBO, the term $U_\mathrm{c}$ is not
big, but the glassy exponent $\mu$ is sizeable, therefore the
second term becomes quite important. In the low temperature limit,
we could use this approximation to extrapolate the curve $T/Q$
down to zero temperature, the intercept gives $U_\mathrm{c}(0)/k_B$.
The relaxation rate $Q$ is related to the balance of the two terms $U_\mathrm{c}$ and $\mu CT$ and shows a complex temperature dependent behavior.

\section{Results and discussion}
\subsection{Transient superconducting current density and second peak effect}

\begin{figure}
\includegraphics[width=8cm]{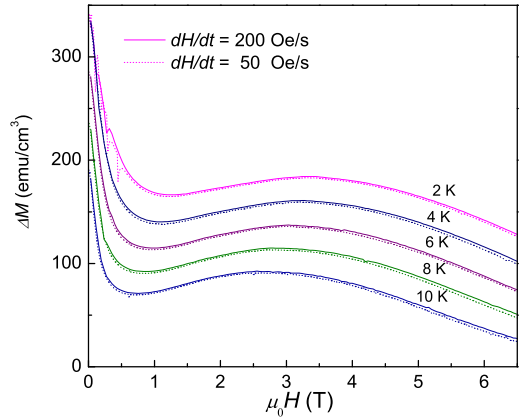}
\caption{(Color online) Field dependence of $\Delta M$ with different
field sweeping rates at various temperatures ranging
from 2 K to 10 K.}\label{fig4}
\end{figure}

\begin{figure}
\includegraphics[width=8cm]{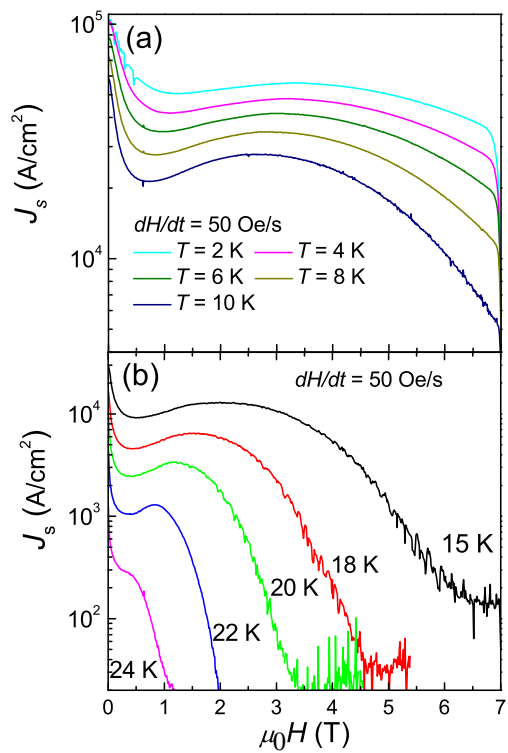}
\caption{(Color online) (a) Magnetic field dependence of the
calculated transient superconducting current density $J_\mathrm{s}$ based
on the Bean critical state model at temperatures ranging from 2 K
to 10 K. (b) Magnetic field dependence of the calculated transient
superconducting current density based on the Bean critical state
model at temperatures ranging from 15 K to 24 K. }\label{fig5}
\end{figure}

In Fig. 4, we show the field dependent $\Delta M$ with different field
sweeping rates of 200 Oe/s and 50 Oe/s respectively at different
temperatures ranging from 2 K to 10 K. Fig. 5 shows the field
dependence of $J_\mathrm{s}$ with the magnetic field sweeping rate of 50
Oe/s by using Eq. (3). The calculated $J_\mathrm{s}$ at 2 K at zero magnetic field can reach
up to 10$^5$ A/cm$^2$, which is an order of magnitude larger than
the values reported in literatures \cite{BARILOphysicac254}, while
the $J_\mathrm{s}$ is much smaller than the cuprate and iron-based
superconductors. As presented in Fig. 5(b), in the high
temperature region $J_\mathrm{s}$ decreases greatly due to severe flux motion.

From the MHL curves in Fig. 3 and $J_\mathrm{s}$-$B$ curve in Fig. 5, we can observe second peak (SP)
effect (fish-tail effect) in a very wide temperature
region ranging from 2 K to 22 K. The second peak is relative to the first one near zero magnetic field in MHL or $J_\mathrm{s}$-$B$ curve, and the magnetic field of the second peak position is defined as $H_\mathrm{sp}$ which is dependent on
temperature. We can see this clearly from Fig. 3 that the peak position of SP moves toward lower magnetic field as the
temperature rises. This feature was also observed in the cuprate
superconductors, such as YBa$_2$Cu$_3$O$_7$ (YBCO) \cite{LKPRB49}, as well as in the
iron-based superconductors\cite{YangHAPL,BSPRB81}. The second peak
effect has been intensively studied previously, and several possible
mechanisms have been proposed. These
include (1) Inhomogeneities in the sample, such as nonuniform
oxygen distribution in the cuprates, which generates
oxygen-deficient regions acting as extra pinning centers in high
magnetic field\cite{MDNATURE}, and thus enhances the
superconducting current density; (2) A crossover from fast
relaxation to slow relaxation, the transition occurs between the
single vortex regime and the collective pinning and creep regimes
with slower relaxation at sufficiently high magnetic fields
\cite{LKPRB49}; (3) A crossover in flux dynamics from elastic to
plastic vortex creep \cite{YAPRL77}. It is well-known that the SP
effect in YBCO and Bi$_2$Sr$_2$CaCu$_2$O$_8$ (Bi2212) samples
exhibits in different ways. In Bi2212, the SP field is low
(usually a few hundreds of oersted) and weakly temperature
dependent, but SP in YBCO occurs at a high magnetic field
(usually a few or more than ten Tesla) and is strongly temperature
dependent. Our present results in BKBO show that it is more like
the SP in YBCO. This probably suggests that the SP in BKBO and
YBCO is due to the similar reasons. One of possible explanations is that
both systems are more three dimensional and containing oxygen
vacancies. The oxygen content may not be very uniform leading to
many local random pinning centers.

\subsection{Magnetization relaxation and its correlation with $J_\mathrm{s}$}

It can be clearly seen from Fig. 4 that the MHL curves demonstrate
a difference in magnitude with different field sweeping rates at a
certain temperature. The larger the field sweeping rate is, the
bigger of the MHL width will be. As we know the magnitude of the
difference between the MHL curves measured at different field
sweeping rates reflects how strong the vortex creep is. In the
cuprate high temperature superconductors, it was found that the
separation between different MHLs is quite
large\cite{WenHHPhysicaC1998}. At high temperatures, however, the
magnitude of the diamagnetic moment greatly decreases with
increasing magnetic field. Based on Eq. (4) for treating the magnetization relaxation, we
calculated the magnetic field and temperature dependence of the
dynamical relaxation rate $Q$ from the data shown in Fig. 4. As presented in Fig.
6(a), it can be seen that $Q$ decreases first and then increases
with increasing magnetic field at temperatures of 2 K and 10 K,
showing a minimum of $Q$ in the intermediate magnetic field
region. This corresponds very well with the width of the MHLs or
the transient superconducting current density qualitatively as
shown in Fig. 6(b). However, not like that reported in
Ba(Fe$_{0.92}$Co$_{0.08}$)$_2$As$_2$
\cite{BSPRB81,KonczykowskiPRB}, we do not observe a clear
crossover of $Q$ value near zero field, which was interpreted as a
crossover from the strong intrinsic pinning near zero field to a
collective pinning at a high magnetic field. This may suggest that
there is just one kind of pinning mechanism for the measured BKBO
single crystal.

As we know that magnetization relaxation rate $Q$ is affected by not
only the vortex pinning but also the interaction between
vortices. In the normal circumstance, the magnetization relaxation
$Q$ will increase with increasing magnetic field because of the
enhanced interaction between vortices at a certain temperature.
The irreversible magnetization or the transient critical current
density $J_\mathrm{s}$ will decrease with the magnetic field as the flux
creep is enhanced. However, in Fig. 6 we can observe
non-monotonic magnetic field dependence of both $Q$ and $J_\mathrm{s}$. Indeed, one can clearly see
that when the MHL is getting wider versus magnetic field, the
relaxation rate is getting smaller, showing a slower magnetization
relaxation, i.e., there is a close correlation between field dependent behaviors of $Q$ and $\Delta M$(or
$J_\mathrm{s}$). This feature is very similar to those in cuprate and
iron based superconductors\cite{YeshurunSP,BSPRB81}. What we want to notice is that
the positions of minimum $Q$ are roughly corresponding to the
location where $\Delta M$ takes the maximum. This also
reminds us that the second peak effect appearing here is just
reflecting a dynamical process of the vortex entity, but not the
detailed characteristics of the pinning centers, because one
cannot guarantee the pining mechanism is the same in all these
different superconductors. It would be very
interesting to measure the magnetization relaxation in a long time
scale, to reveal whether the SP effect corresponds to a static
vortex phase transition\cite{Yeshurun2}.

\begin{figure}
\includegraphics[width=8cm]{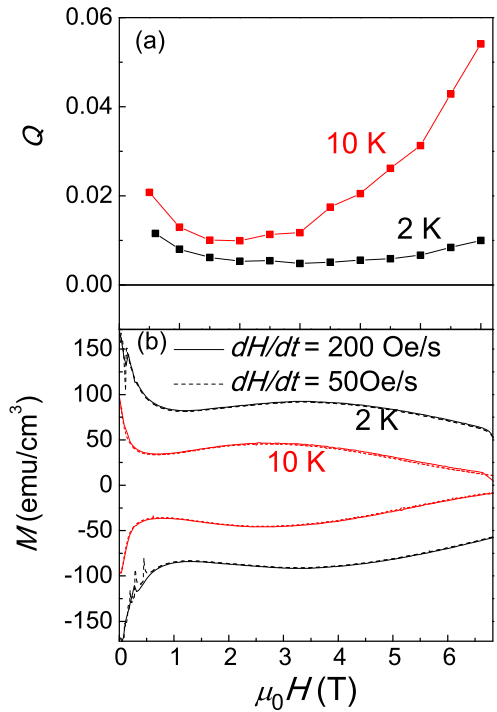}
\caption{(Color online)(a) The magnetic field dependence of
magnetization-relaxation rate $Q$ at temperatures of 2 K and 10 K
obtained from curves in (b) with Eq.(4). (b) The MHL curves at
temperatures of 2 K and 10 K. Solid lines stand for the magnetization measured with a magnetic
field sweeping rate of 200 Oe/s while the dashed line to 50 Oe/s. Several sharp peaks of magnetization in the field ascending process below 0.5 T at
2K are induced by the flux jump effect.}\label{fig6}
\end{figure}

\begin{figure}
\includegraphics[width=8cm]{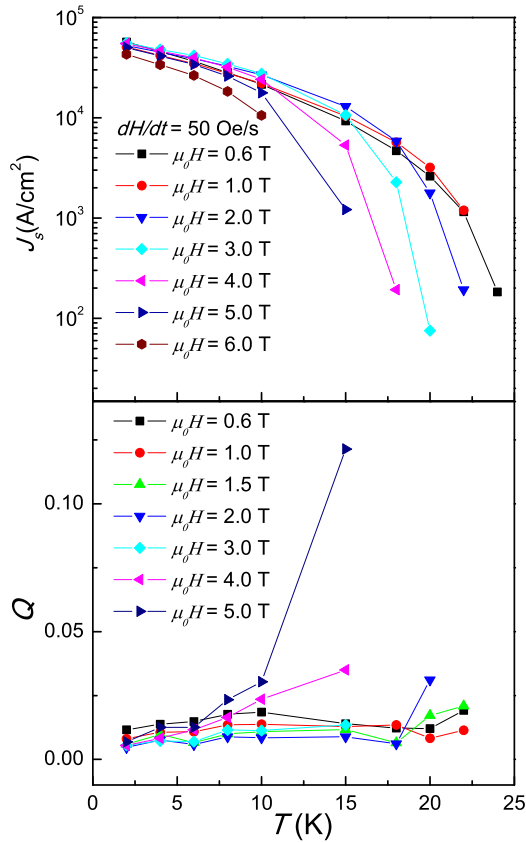}
\caption{(Color online) (a) Temperature dependence of log
$J_\mathrm{s}$ at different magnetic fields ranging from 0.6 T to
6 T, the data is the same as that presented in Fig. 5. (b) Temperature
dependence of magnetization relaxation rate $Q$ at various
magnetic fields from 0.6 T to 5.0 T obtained from the
corresponding curves in Fig. 3 with Eq.(4).}\label{fig7}
\end{figure}

Fig. 7 shows the temperature dependence of transient
superconducting current density $J_\mathrm{s}$ and dynamical relaxation
rate $Q$ taken from field dependent values at various temperatures, respectively.
We notice that in the low and intermediate temperature region, the curves of $\log J_\mathrm{s}(T)$
at different magnetic fields almost merge together below 6 T, which disperses clearly in high temperature region.
This behavior is similar to that of Ba(Fe$_{0.88}$Co$_{0.12}$)$_2$As$_2$\cite{BSPRB81}, and may be
caused by the SP effect which prevents the rapid decreasing of $J_\mathrm{s}$. Associated with the SP effect, the magnetization
relaxation rate $Q$ is inversely related to transient
superconducting current density $J_\mathrm{s}$ as shown in Fig. 7(b).
As we mentioned already that there is a plateau in $Q - T$ curve below
18 K, and below 2 T the $Q$ value decreases with increasing
field, all these features demonstrate that the $Q$-plateau and the
SP effect are closely related. When it is in the collective creep
regime, the magnetization relaxation rate is low and exhibits a
plateau, thus the transient critical current density $J_\mathrm{s}$ decays
slowly with temperature. It seems that all these features can be
explained coherently.

As addressed above, a plateau of $Q$ appears
in the intermediate temperature region which is followed by a severe increase in the high temperature region. This behavior of
magnetization relaxation rate was also observed in cuprate
superconductor YBCO\cite{RGprl72} and
iron-based superconductors, such as
Ba(Fe$_{0.92}$Co$_{0.08}$)$_2$As$_2$ \cite{BSPRB81} and
SmFeAsO$_{0.9}$F$_{0.1}$ \cite{YangHPRB}. This plateau cannot be
understood within the picture of single vortex creep with a rigid
hopping length as predicted by the Kim-Anderson model, but perhaps
due to the effect of collective flux creep. The reason is that, in the Kim-Anderson model,it is quite easy to derive that $T/Q(T)=U_\mathrm{c}(T)$. Suppose $U_\mathrm{c}(T)$ is a weak temperature dependent
function, we have $Q(T)\propto T$, which contradicts the observation of a $Q$-plateau in the intermediate temperature region. As described in Eq. (5), the relaxation rate $Q$ is dependent on both $U_\mathrm{c}(T)$ and $\mu CT$. In the intermediate temperature region, the term of $\mu CT$ becomes much larger than $U_\mathrm{c}/k_B$ and we will get almost constant value of $Q$, which provides a simple explanation of the plateau.

A coherent picture to interpret the complex temperature and magnetic field dependence of $Q$ and $J_s$ is thus proposed in the following way. When the magnetic field increases from zero, the
vortex system runs into the vortex glass regime with much smaller
relaxation rate. In the high field region, the magnetic relaxation
rate $Q$ goes up drastically, meanwhile the transient
superconducting current $J_\mathrm{s}$ drops down
quickly as shown in Fig. 7(a), which could be interpreted as a crossover from the
elastic motion to a plastic one of the vortex system.

\subsection{Analysis based on the collective pinning/creep model}

\begin{figure}
\includegraphics[width=8cm]{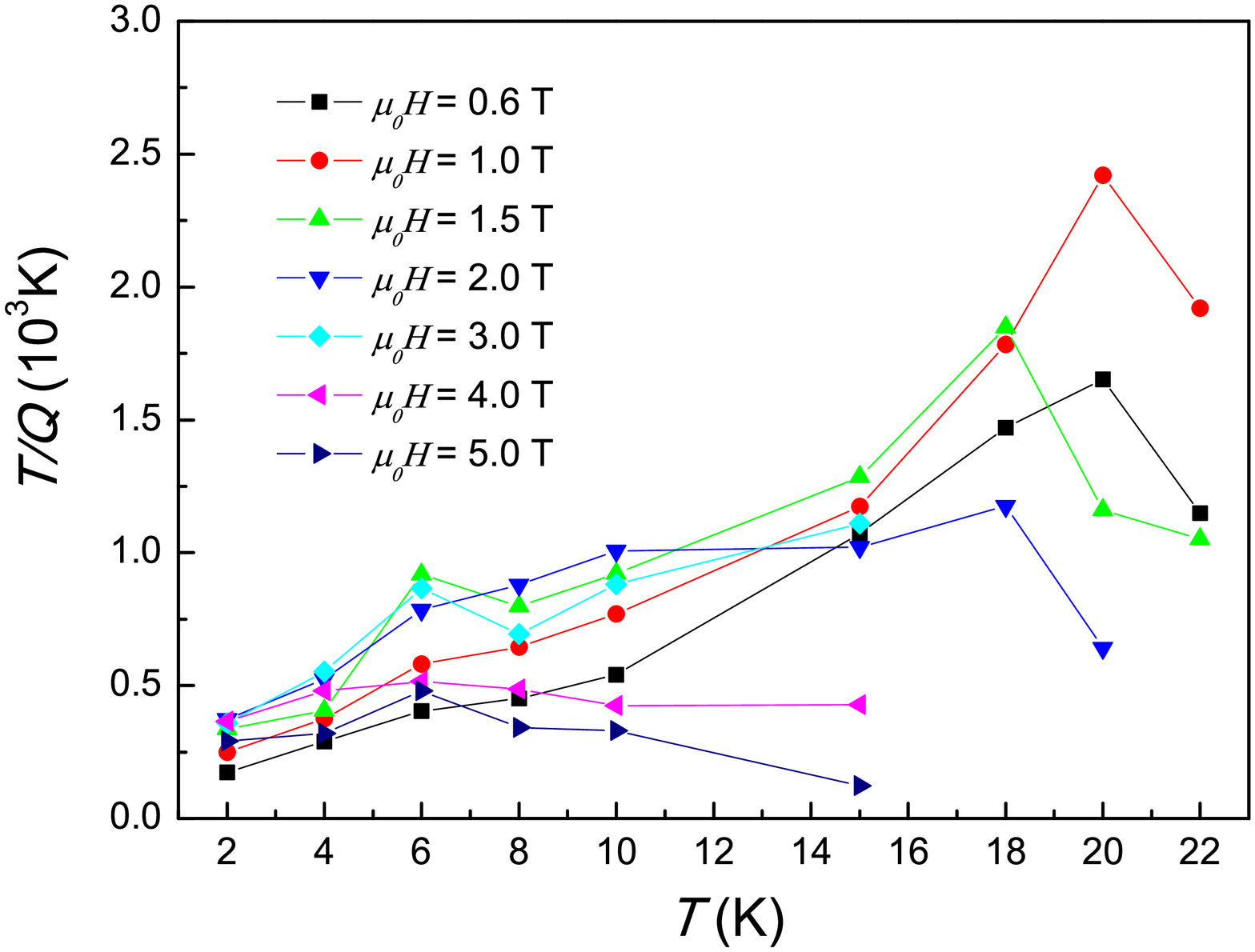}
\caption{(Color online) Temperature dependence of the ratio $T/Q$
at different magnetic fields ranging from 0.6 T to 5.0
T.}\label{fig8}
\end{figure}

Now let's have a quantitative consideration on the vortex dynamics based on
the collective pinning/creep model. According to Eq. (5), the
intercept of $T/Q$-$T$ curve will give $U_\mathrm{c}/k_B$ and the
slope gives rise to $\mu C$ if we assume $U_\mathrm{c}$ and $\mu$
is weakly temperature dependent. In Fig.8 we present the $T/Q$ vs.
$T$ at different magnetic fields ranging from 0.6 T to 5 T. One
can see that the intercept $U_\mathrm{c}/k_B$  is actually very
small, about 111 K at 1 T, and 198 K at 2 T, which is much smaller
than the value of over 3000 K in MgB$_2$ \cite{MgB2Uc} and also
smaller than about 300-500 K in YBCO \cite{WenHHPhysicaC95,WENJAC}.
However, since $J_\mathrm{s}$ is quite large and the relaxation
rate is very small, the collective pinning should play the role
here. As we discussed above, in the model of collective creep, the
glassy exponent $\mu$ becomes influential on the vortex dynamics.
From Eq.(5), one can see that $\mu C$ can be determined from the
slope of $T/Q$ vs $T$. From our data, we obtained the value $\mu C
= 63$ in the low temperature regime at 1 T from Fig. 8. So in the
intermediate temperature region, $U_\mathrm{c}/k_B\ll\mu C$ and we
will get the plateau of $Q\approx1/\mu C$ as shown in Fig. 7(b). Meanwhile,
the parameter $C$ can be determined by the slope of $-d\ln
J_\mathrm{s}/dT$ vs $Q/T$\cite{WenHHPRL1997,WenHHPhysicaC95}, and
we get the $C=24.8$ and the glassy exponent $\mu\approx 2.54$. The value of $\mu$ is much larger than
1 and clearly shows a collective creep effect. Actually, $\mu$
can also be estimated\cite{WenHHPRL1997} from $\mu=-Qd\ln^2J_\mathrm{s}/d\ln E^2$.
From this equation, one can imagine that large $\mu$ means a stronger glassy effect, showing a
strong downward curvature in the $\ln E$ vs $\ln J_\mathrm{s}$ curve. All these
indicate that the vortex pinning and dynamics in BKBO can be
described by the collective pinning/creep model with weak
characteristic pinning energy $U_\mathrm{c}$ but large glassy exponent
$\mu$.

\subsection{Characteristics of pinning force density}

\begin{figure}
\includegraphics[width=8cm]{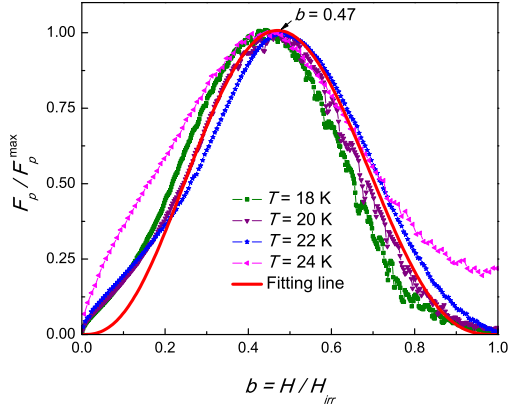}
\caption{(Color online) The scaling of normalized pinning force
density $F_p/F_p^\mathrm{max}$ as a function of the reduced field $b=H/H_\mathrm{irr}$. Curves at different temperatures are almost scaled together. A fitting result with Eq. (6)
is presented as the red line, and the maximum locates at $b\approx 0.47$. }.\label{fig9}
\end{figure}

In order to get further insight into the origin of the second peak
effect, we need to analyze the pinning force density $F_p$ which
is proportional to $J_\mathrm{s} H$. In Fig. 9 we show the $F_
\mathrm{p}$ normalized to its maximum value as a function of
reduced field $b=H/H_\mathrm{irr}$, and $H_\mathrm{irr}$ is the
irreversible magnetic field determined using a criterion of
$J_\mathrm{s}= 20$ A/cm$^2$. Although there is uncertainty to
determine $H_\mathrm{irr}$, the curves of normalized pinning force
at different temperatures seem to scale well and have the similar
maximum value, which is different from the poor scaling results
observed in the thick films of BKBO\cite{BKBOfilm}. The maximum of
the pinning force density locates at $b\approx{0.47}$. We use the
following expression to study the pinning mechanism for the sample
Ba$_{0.66}$K$_{0.32}$BiO$_{3+\delta}$
\begin{equation}
\frac{F_\mathrm{p}}{F_p^\mathrm{max}}=A b^p(1-b)^q,
\end{equation}
where $A$, $p$ and $q$ are the parameters, and the values of $p$
and $q$ can tell us the characteristic properties of the vortex
pinning mechanism in the sample. We use this equation to fit the
data, which is presented as the red line in Fig. 9. It can be seen
that the fitting curve with $p=2.79$ and $q=3.14$ catches the main
feature of the experimental data. As we know the cuprate
high-temperature superconductors are usually satisfied with the
relation $q > p$, e.g., $q=4$, $p=2$ were obtained for YBCO single
crystals \cite{JNPHYSICAC169}, and $q=2$, $p=0.5$ for BSCCO single
crystal \cite{RosePHYSICAC170}. In this sample, $b$ for the
maximum value of the pinning force density is near 0.5 which
corresponds to the $\delta\kappa$-type pinning\cite{KRAMERJAP44}, however this kind of
pinning requests $p=q=1$. Further studies are still required to
resolve this discrepancy.

\section{Vortex phase and general discussion}

\begin{figure}
\includegraphics[width=8cm]{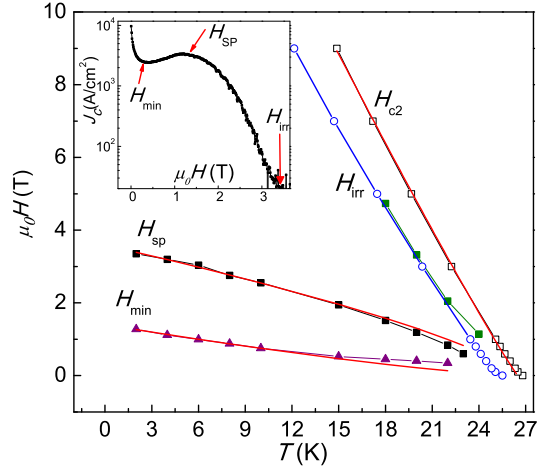}
\caption{(Color online) The phase diagram of the BKBO sample. The open symbols are taken from the resistive measurements shown in Fig. 2, while the solid ones are taken from the $J_s$-$\mu_0H$ curve in Fig. 5. The inset shows a typical example for how to determine characteristic
fields $H_\mathrm{irr}$, $H_\mathrm{sp}$ and $H_\mathrm{min}$. All red lines in this figure are the fitting curves with the formula
$H(T) = H(0)(1-T/T_\mathrm{c})^n$. For the irreversibility field $H_\mathrm{irr}(T)$, the open circles correspond to the data determined from resistivity, the filled green squares represent the data determined from the irreversible magnetization.}\label{fig10}
\end{figure}

In Fig. 10, we present the vortex phase diagram of the sample
Ba$_{0.66}$K$_{0.32}$BiO$_{3+\delta}$. The upper critical field
$H_\mathrm{c2}$ and the irreversible field $H_\mathrm{irr}$ shown
as open symbols in Fig. 10 are determined by a criterion of
$90\%\rho_\mathrm{n}$ and $10\%\rho_\mathrm{n}$ in Fig. 2. The
three characteristic fields are shown as solid symbols in Fig. 10,
i.e., the second magnetization peak field $H_\mathrm{sp}$,
$H_\mathrm{min}$ determined at the minimum point of $\Delta M$ or
$J_s$ between the first and the second magnetization peak, and the
irreversibility field $H_\mathrm{irr}$ determined from the field
dependent $J_\mathrm{s}$ curve in Fig. 5. In order to get more
information, we use the expression $H(T) =
H(0)(1-T/T_\mathrm{c})^n$ to fit these curves. We got the
following values: $n=1.32$ for $H_\mathrm{min}$ and
$H_\mathrm{min}(0)=1.4$ T, $n=0.723$ for $H_\mathrm{sp}$ and
$H_\mathrm{sp}(0)=3.59$ T, $n=1.07$ for $H_\mathrm{c2}$ and
$H_\mathrm{c2}(0)=21.6$ T. The exponent $n$ from the
$H_\mathrm{sp}$-$T$ curve is almost half less than that of YBCO
\cite{LKPRB49}. It is clear that in BKBO, the separation between
$H_\mathrm{irr}$-$T$ and $H_\mathrm{c2}$-$T$ is small, which is
similar to YBCO and indicates a rather weak vortex fluctuation. In
addition, one can see that the separation between
$H_\mathrm{sp}$-$T$ and $H_\mathrm{irr}$-$T$ curves is quite
large. One cannot interpret this large region as due to the
non-uniform distribution of disorders or pinning centers. However,
it is reasonable to understand this region as the plastic flux
motion since this phase can gradually ``melt'' through losing the
rigidity of the vortex manifold. It is interesting to note that
the $H_\mathrm{sp}(T)$ looks very similar to that in YBCO, but
very different from that in Bi2212. Therefore we believe that the
second peak effect, at least in BKBO and YBCO, may be induced by
the similar reason. It is quite possible that the elastic energy
which depends on the shear module $C_{66}$ is an influential
factor for the occurrence of the second peak effect.

\section{Concluding Remarks}
We have investigated the vortex dynamics and phase diagram through measuring the magnetization and its relaxation
on a Ba$_{0.66}$K$_{0.32}$BiO$_{3+\delta}$ single
crystal with transition temperature $T_\mathrm{c}=27$ K. Second magnetization peak has been observed in wide
temperature region from 2 K to 24 K. It is found that through out the non-monotonic magnetic field dependence of the magnetization, the
relaxation rate is inversely related to the transient critical
current density, indicating that the SP effect is dynamical in
nature.  It is found that many
observed features can be coherently described by the collective
pinning/creep model. Through the fitting and analysis, we find
that the characteristic pinning energy $U_\mathrm{c}$ is quite small (about
198 K at 2 T), but the glassy exponent $\mu$ is quite large, which
induces a relatively small magnetization relaxation rate. A universal scaling law for the pinning force density $F_\mathrm{p}/F_\mathrm{p}^\mathrm{max}$ vs $H/H_\mathrm{irr}$ is found, which suggests that the pinning mechanism is probably $\delta \kappa$-type. The
characteristics of the SP effect and magnetization relaxation as
well as the vortex dynamics of the system allow us to conclude
that it is more like those in the cuprate superconductor YBCO.

\section{Acknowledgments}
We thank
Xiang Ma and Dong Sheng for the assistance in SEM/EDS
measurements. This work was supported by NSF of China with the project numbers 11034011 and 11190020, the
Ministry of Science and Technology of China (973 projects:
2011CBA00102, 2012CB821403) and PAPD.

\end{document}